\documentclass[superscriptaddress,nobibnotes,amsmath,amssymb,notitlepage,twocolumn,prl]{revtex4-2}

\usepackage{pdfpages}
\usepackage{etoolbox} 

\usepackage{mathptmx}
\usepackage{braket}
\usepackage{graphicx}
\usepackage[caption=false]{subfig}
\usepackage{hyperref}
\hypersetup{colorlinks=true, linkcolor=blue, citecolor=blue, urlcolor=blue}

\graphicspath{{./img/}}

\newcommand{\beq}[1]{\begin{equation}\label{#1}}
\newcommand{\eep}{\;.\end{equation}}
\newcommand{\eec}{\;,\end{equation}}
\newcommand{\eeq}{\end{equation}}

\newcommand*\dd{\mathop{}\!\mathrm{d}} 

\DeclareMathOperator{\tr}{tr}

\newcommand{\lb}{\left(}
\newcommand{\rb}{\right)}

\newcommand*{\heading}[1]{\belowpdfbookmark{#1}{#1}{\textit{#1.---}}\ignorespaces}
\let\section\heading 

\newcommand*\chem[1]{\ensuremath{\mathrm{#1}}} 

\renewcommand{\a}{\alpha}
\renewcommand{\b}{\beta}
\renewcommand{\d}{\delta}
\newcommand{\ep}{\epsilon}
\renewcommand{\k}{\kappa}
\newcommand{\la}{\lambda}
\renewcommand{\th}{\theta}
\newcommand{\p}{\phi}

\newcommand{\D}{\Delta}
\newcommand{\Om}{\Omega}

\DeclareMathAlphabet{\mathcal}{OMS}{cmsy}{m}{n} 

\newcommand{\Df}{\mathcal{D}}   
\newcommand{\Ef}{\mathcal{E}}   

\renewcommand{\vec}[1]{{\bf #1}}

\newcommand{\x}{\vec{x}}
\renewcommand{\v}{\vec{v}}
\newcommand{\kv}{\vec{k}}
\newcommand{\av}{\vec{a}}

\renewcommand{\P}{\vec{P}}

\newcommand{\ZA}{Z^{\rm A}}
\newcommand{\ZS}{Z^{\rm S}}

\begin{document}

\makeatletter
\patchcmd{\@outputpage@head}{\@ifx{\LS@rot\@undefined}{}{\LS@rot}}{}{}{}
\makeatother

\title{Asymmetric dynamical charges in two-dimensional ferroelectrics}


\newcommand{\Liege}{Theoretical Materials Physics, Q-MAT, University of Liège, B-4000 Sart-Tilman, Belgium}
\newcommand{\HarvardSeas}{John A.~Paulson School of Engineering and Applied Sciences, Harvard University, Cambridge, Massachusetts 02138, USA}

\author{Daniel Bennett}
\email{dbennett@seas.harvard.edu}
\affiliation{\Liege}
\affiliation{\HarvardSeas}

\author{Philippe Ghosez}
\affiliation{\Liege}

\begin{abstract}
Ferroelectricity is commonly understood in terms of dynamical charges, which represent the dipole moments generated by atomic displacements or the forces induced by electric fields. 
In ferroelectrics with a high degree of symmetry, the dynamical charges are typically symmetric tensors, and can be visualized as ellipsoids. 
In van der Waals (vdW) materials which break centrosymmetry, a new type of ferroelectricity arises which differs greatly from conventional ferroelectrics.
The polarization is purely electronic, arising from an interlayer charge transfer, and most of the polarization generated is perpendicular to atomic motion.
We show that the unconventional properties of vdW ferroelectrics are manifested in their dynamical charges, which exhibit spatial modulation and intrinsic asymmetry. 
Dynamical charges in vdW ferroelectrics, and more generally, any strongly anisotropic ferroelectric, can be visualized as deformable, non-ideal ellipsoids dependent on the atomic configuration. 
Furthermore, we show that, due to the mixed electrostatic boundary conditions employed for two-dimensional materials, non-diagonal dynamical charges in 2D materials are always asymmetric.
\end{abstract}

\maketitle

\section{Introduction}
Ferroelectricity, a spontaneous polarization that can be switched with an applied electric field, is a material property not only of great interest in terms of fundamental physics, but also enables many useful applications in semiconductor technologies \cite{scott2007applications}.
Ferroelectricity arises from the softening of unstable polar phonon modes in a non-polar reference state, allowing the material to relax into one of several lower-energy polar states characterized by a spontaneous polarization. 
This polarization can be reversibly switched between states using an applied electric field.
For example, in oxide perovskites ABO$_3$, a commonly studied family of `conventional' ferroelectrics, the polarization is understood to originate from polar phonon mode displacements, typically the off-centering of the B cation with respect to the oxygen octahedra \cite{cohen1992origin}.

In conventional ferroelectrics, the polarization $\P$ is well-described by
\beq{P-x}
P_{\b} = Z^{*}_{\k,\a\b}x_{\k,\a}, \quad
Z^*_{\k,\a\b} = \Om\frac{\partial P_{\b}}{\partial x_{\k,\a}} = \frac{\partial F_{\k,\a}}{\partial \Ef_{\b}}
\eec
i.e.~the product of the polar mode displacements $\x_{\k}$ of atoms $\k$, and the dynamical charges $Z^*$, where $\Om$ is the unit cell volume.
The dynamical charges are mixed derivatives of the free energy, and can either be thought of as the dipole generated by an atomic displacement, or the forces $\vec{F}_{\k}$ induced by an electric field $\Ef$ \cite{gonze1997dynamical,ghosez1998dynamical}.
The dynamical charges can be calculated as a linear response \cite{baroni1987green} using density functional perturbation theory (DFPT) \cite{gonze1995adiabatic,ghosez1998dynamical,gonze1997dynamical,baroni2001phonons,Wu2005}, and have proven to be a useful concept for understanding the electrical properties of insulators:
they provide an intuitive way to visualize charges in materials in a dynamical sense, which can differ significantly from the static charges, by taking into consideration the complex dielectric environment surrounding the ions.
For a crystal in a phase with a high degree of symmetry, the antisymmetric part of $Z^*$ is zero or negligible, it has real eigenvalues as well as orthogonal eigenvectors. 
This allows the dynamical charges to be visualized as ellipsoids: $\sum_{i=1}^3 \la_i x_i^2 \leq 1$ \cite{ghosez1998dynamical}, where $x_i$ are coordinates along the eigenvectors $\v_i$, which define the principle axes, and $\la_i$ are the eigenvalues.
The dynamical charges are also a well-defined quantity, being a derivative of the total polarization, in contrast to polarization itself, which in a periodic crystal is a lattice-valued quantity and must be treated using the modern theory of polarization \cite{king1993theory,vanderbilt1993electric}.
They also have the powerful advantage that they allow the polarization to be decomposed into contributions from the displacements of individual ions, and displacements in individual directions.
Additionally, the dynamical charges are essential for describing the relaxed-ion responses of materials, such as dielectric permittivity, piezoelectricity and optical responses \cite{gonze1997dynamical,bennett2024electrostriction}.

\begin{figure}[t!]
\centering
\includegraphics[width=\linewidth]{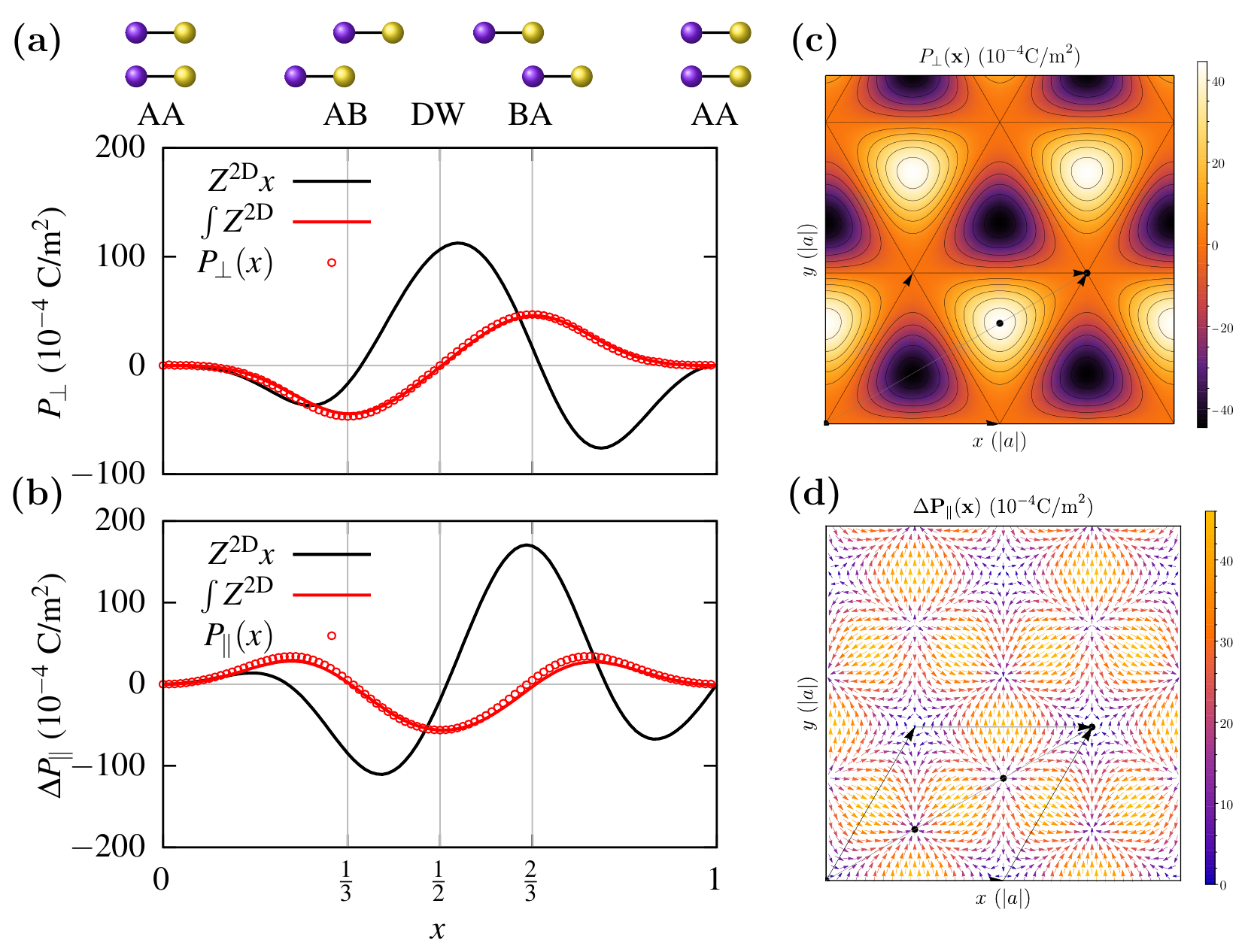}
\caption{
{\bf (a)} Out-of-plane and 
{\bf (b)} in-plane polarization as a function of relative displacement along the unit cell diagonal, obtained from Berry phases (red dots), 
integrating the dynamical charges (red line) and multiplying the dynamical charges by the relative displacement (black line).
The in-plane polarization is shown as a projection onto the unit cell diagonal: $P_{\parallel} = \P_{\parallel}\cdot \frac{1}{\sqrt{2}}\lb \av_1 + \av_2\rb$.
The Berry phases were calculated with mixed electrostatic boundary conditions: $\{\Ef_{\parallel},\Df_{\perp}\}=0$.
The dynamical charges for the same boundary conditions, $Z^{\rm 2D}$, were obtained by modifying the Born effective charges as described by Eq.~\eqref{eq:Z-mix}.
The high symmetry stackings AA ($x=0$), AB ($x=\frac{1}{3}$), DW ($x=\frac{1}{2}$) and BA ($x=\frac{2}{3}$) are indicated by the ticks and sketched above.
{\bf (c)} Out-of-plane and 
{\bf (d)} in-plane polarization as a function of relative stacking in 2D. 
A primitive cell of commensurate bilayer hBN is sketched.
}
\label{Fig1}
\end{figure}

\begin{figure*}[t]
\centering
\includegraphics[width=\linewidth]{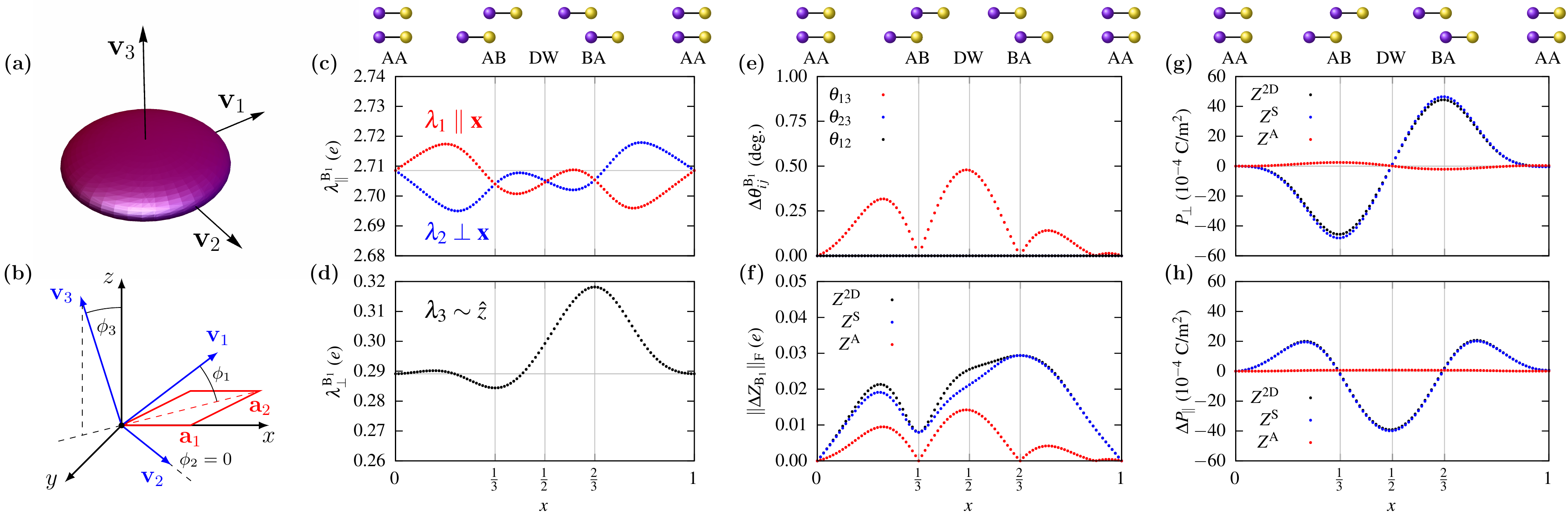}
\caption{
{\bf (a)} Sketch of a flat ellipsoid, representing the dynamical charges in vdW ferroelectrics.
{\bf (b)} Illustration of eigenvectors  $\v_i$ and their deflections with respect to principle axes $\p_i$. 
A primitive cell of hBN is sketched in red, with lattice vectors $\av_1 = \hat{\x}$ and $\av_2 = \frac{1}{2}\hat{\x} + \frac{\sqrt{3}}{2}\hat{\vec{y}}$.
The direction of sliding along the unit cell diagonal is indicated by the red dashed line.
Eigenvalues {\bf (c)} in-plane and {\bf (d)} out-of-plane, for the B atom in the bottom layer. The eigenvalues of all atoms are available in Ref.~\cite{SM}.
{\bf (e)} Deviation from orthogonality of the eigenvectors, ${\D\th_{ij} = \left|\cos^{-1}{\lb \v_i\cdot\v_j\rb} - \frac{\pi}{2}\right|}$.
{\bf (f)} Frobenius norm of the change in $Z^{\rm 2D}$ (black), and symmetric (red) and antisymmetric (blue) parts, as a function of $\x$: $\|\D Z\|_{\rm F} \equiv \|Z(\x)-Z(0)\|_{\rm F}$.
{\bf (g)} Out-of-plane and {\bf (h)} in-plane polarization, obtained by integrating $Z^{\rm 2D}$ (black), and by integrating the symmetric (blue) and antisymmetric (red) parts individually.
}
\label{Fig2}
\end{figure*}

Recently, a new type of ferroelectricity in two-dimensional (2D) layered materials was proposed \cite{li2017binary} and experimentally observed \cite{zheng2020unconventional,stern2020interfacial,yasuda2021stacking,wang2022interfacial,weston2022interfacial,ko2023operando,molino2023ferroelectric,van2024engineering}. 
In layered van der Waals (vdW) systems such as hexagonal boron nitride (hBN) or transition metal dichalcogenides (TMDs), where centrosymmetry is broken (typically by artificial stacking-engineering), an out-of-plane polarization $P_{\perp}$, which is purely electronic, occurs via an interlayer charge transfer, the magnitude of which is determined by the relative stacking between the layers, see Fig.~\ref{Fig1}.
$P_{\perp}$ can be switched by a relative sliding of one third of a unit cell diagonal between the layers, known as van der Waals sliding \cite{stern2020interfacial}, which results in `vdW ferroelectricity'.
This new and unconventional type of ferroelectricity has been shown to out-perform state-of-the-art ferroelectric field transistors (FeFET) \cite{yasuda2021stacking},
with a spontaneous polarization that persists at room temperature, retention times of up to one month, and ultrafast switching times as low as 1 ns, which can be achieved using electric field pulses.
In addition, little to no fatigue has been observed up to $10^{11}$ cycles; due to the long-range nature of the interlayer electrostatic interactions, no bonds are broken as one layer slides over the other to invert the polarization.

When there is a relative twist or lattice mismatch between the layers, forming a moir\'e superlattice, the interlayer charge transfer results in an out-of-plane polarization texture \cite{bennett2022electrically,bennett2022theory}, and the stacking domains which form can be identified as moir\'e polar domains (MPDs), see Fig.~\ref{Fig1} (c), which have been experimentally shown to result in ferroelectricity \cite{yasuda2021stacking,ko2023operando} via the growing and shrinking of the MPDs in response to an applied field \cite{bennett2022electrically,bennett2022theory}.
Additionally, it was recently proposed that the different relative stackings also give rise to an in-plane polarization $\P_{\parallel}$ \cite{bennett2023polar,bennett2023theory}, see Fig.~\ref{Fig1} (d), which in moir\'e superlattices result in topologically nontrivial polarization textures. 
This makes vdW materials, both twisted and untwisted, a promising avenue for engineering ferroelectrics on the nanoscale.

Although the origin is still based on symmetry breaking, unstable polar modes (shear modes for vdW sliding) and charge transfer, the mechanism for ferroelectricity in vdW materials is very unique when compared to conventional ferroelectrics such as oxide perovskites.
First, in conventional ferroelectrics the polarization is typically parallel to the atomic displacements: $\P \parallel \x$, but in vdW ferroelectrics a significant polarization is generated {\it perpendicular} to the atomic motion, i.e.~an out-of-plane polarization $P_{\perp}$ is generated in response to in-plane sliding $\x$ \cite{bennett2023polar,bennett2023theory}.
Second, in conventional ferroelectrics, the dynamical charges are approximately constant for small displacements, and the polarization is given by Eq.~\eqref{P-x}.
In vdW ferroelectrics however, the Born charges are nonlinear functions of the atomic displacements, and Eq.~\eqref{P-x} is not valid; 
the polarization must be obtained by integrating the dynamical charges \cite{bennett2023polar,bennett2023theory},
\beq{eq:P-int}
P_{\a}(\x) = \int_0^{\x} Z^{*}_{\k,\a\b}(\x')\dd x'_{\k,\b}
\eec
from a non-polar reference configuration to a general configuration $\x$, see Figs.~\ref{Fig1} (a) and (b).

In this letter, we show that the unconventional nature of 2D layered ferroelectrics results in peculiar features of the dynamical charges, using first-principles calculations for rhombohedral bilayer hBN, the prototypical vdW ferroelectric.
Due to the polarization generated perpendicular to atomic motion, the dynamical charges are asymmetric for low-symmetry stackings.
In many common ferroelectrics, the dynamical charges are symmetric or have a negligible antisymmetric part, and can be visualized as regular ellipsoids. 
For systems with a low-degree of symmetry, the dynamical charges cannot be expected to be symmetric, as was reported for \chem{La_2NiMnO_6}, for example \cite{das2009theoretical}.
However, the origin, implications and physical interpretation of asymmetric dynamical charges are not well-known.
We show that when considering mixed electrostatic boundary conditions $\{\Ef_{\parallel},\Df_{\perp}\}=0$ ($\Ef$ is the electric field and $\Df$ is the displacement field), appropriate for 2D materials, the dynamical charges are naturally asymmetric, provided they are non-diagonal.
Furthermore, we show that when the dynamical charges have a small antisymmetric part, they still have real eigenvalues, but the eigenvectors are no longer orthogonal.
In vdW ferroelectrics, the dynamical charges, their eigenvalues and the angles between the eigenvectors modulate as a function of relative stacking between the layers.
Thus, the dynamical charges in vdW ferroelectrics, and more generally in any anisotropic system, can be visualized as deformed ellipsoids which are sensitive to the atomic configurations.
Finally, we discuss the possible consequences of asymmetric dynamical charges.

\begin{figure}[h!]
\centering
\includegraphics[width=\linewidth]{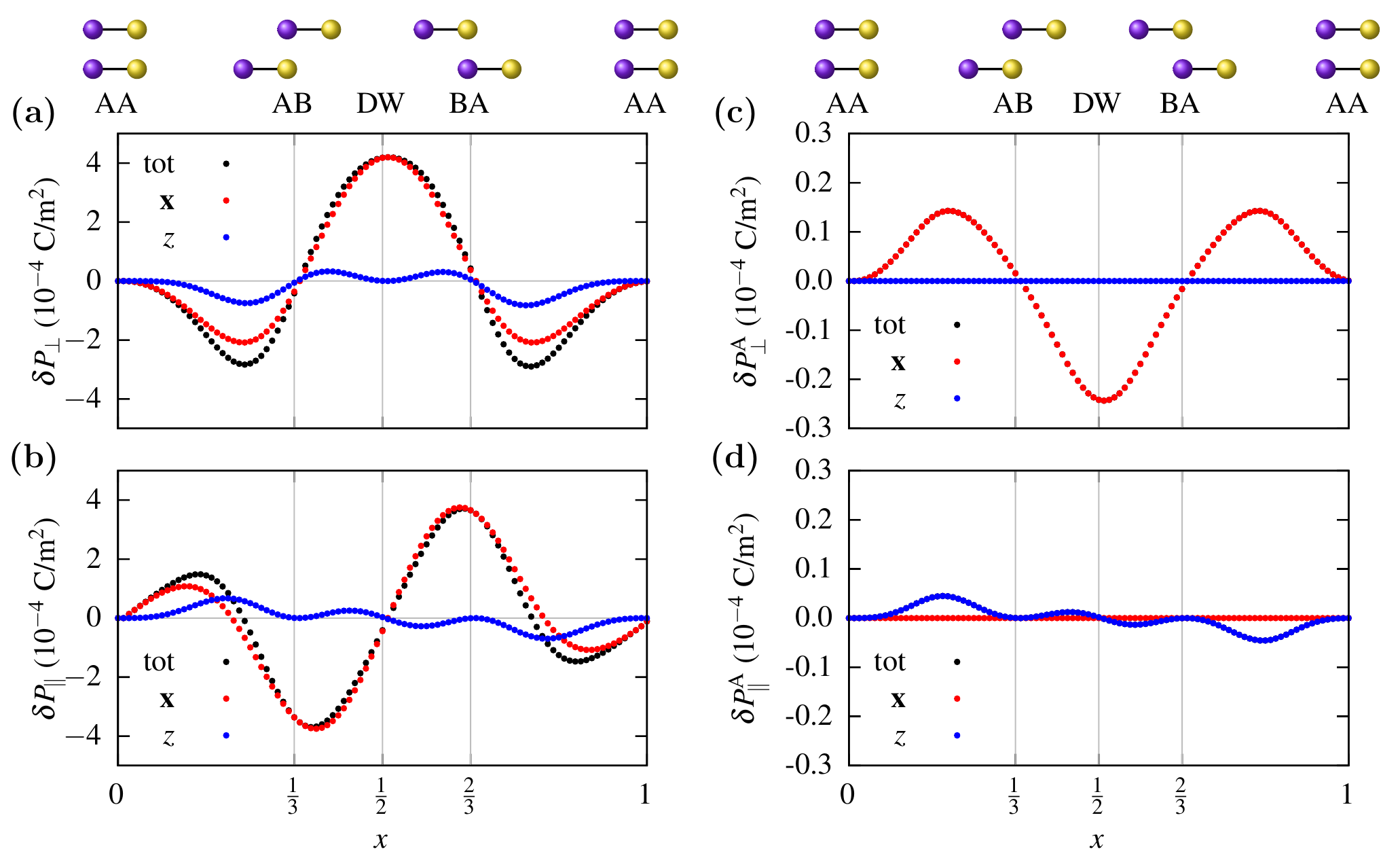}
\caption{
{\bf (a)} Out-of-plane and {\bf (b)} in-plane change in polarization $\d P_{\b} = Z^{\rm 2D}_{\k,\a\b}\d x_{\k,\a}$ (black) and contributions from in-plane motion (red) and out-of-plane corrugation (blue).
{\bf (c)} Out-of-plane and {\bf (d)} in-plane change in antisymmetric polarization $\d P^{\rm A}_{\b} = \ZA_{\k,\a\b}\d x_{\k,\a}$ (black) and contributions from in-plane motion (red) and out-of-plane corrugation (blue).
}
\label{Fig3}
\end{figure}

\section{Results}
First-principles density functional theory (DFT) calculations were performed to simulate bilayer hBN, in the rhombohedral (aligned) stacking, using the {\sc abinit} \cite{gonze2016,gonze2020} code.
Norm-conserving \cite{norm_conserving} {\sc psml} \cite{psml} pseudopotentials were used, obtained from Pseudo-Dojo \cite{pseudodojo}. 
{\sc abinit} employs a plane-wave basis set, which was determined using a kinetic energy cutoff of $1000$ eV. 
A Monkhorst-Pack $k$-point grid \cite{mp} of $16 \times 16 \times 1$ was used to sample the Brillouin zone. 
The revPBE exchange-correlation functional was used \cite{zhang1998comment}, and the vdw-DFT-D3(BJ) \cite{becke2006simple} correction was used to treat the vdW interactions between the layers.

In order to sample the relative stackings between the layers in `configuration space' \cite{carr2018relaxation},
the top layer was translated along the unit cell diagonal over the bottom layer, which was held fixed. 
A fine sampling of 96 points was used, which explicitly includes the high symmetry stackings: the AA stacking, where the two layers are perfectly aligned, the AB and BA where the opposite atoms in neighboring layers are vertically aligned, given by a relative shift of $x=\frac{1}{3}$ or $\frac{2}{3}$ of a unit cell diagonal, respectively, and domain wall (DW) stacking, given by a shift of $x=\frac{1}{2}$ of a unit cell diagonal.
At each point a geometry relaxation was performed to obtain the equilibrium layer separation, while keeping the in-plane atomic positions fixed. 
DFPT calculations were then performed using both electric field and phonon perturbations to calculate the dynamical charges \cite{gonze1997dynamical,ghosez1998dynamical}:
\beq{Z-mixed}
Z^{*}_{\k,\a\b} = \frac{-2ief\Om}{(2\pi)^3}\sum_{n}^{\rm occ} \oint_{\rm BZ} \braket{\partial_{x_{\k,\a}}u_{n,\kv}|\partial_{k_{\b}}u_{n,\kv}}\dd\kv
\eec
where $\ket{u_{n,\kv}}$ are the Bloch states, $\Om$ is the unit cell volume and $f$ is the occupation factor (2 for spin-degenerate systems).

Due to the semi-periodic nature of 2D systems, mixed electrostatic boundary conditions should be employed to calculate electromechanical properties, taking the electric field to be zero in-plane and the displacement field to be zero out-of-plane: $\{\Ef_{\parallel},\Df_{\perp}\}=0$ \cite{royo2021exact}.
The polarization as a function of stacking in Fig.~\ref{Fig1} was calculated using these mixed boundary conditions, 
and differs in the out-of-plane direction by a factor $\ep^{\infty}_{zz}$, 
$\ep^{\infty}$ being the electronic (clamped-ion) permittivity, with respect to calculations performed at fixed electric field in all directions \cite{SM}.
The dynamical charges evaluated at fixed electric field are the well-known Born effective charges $\lb Z^{*}\rb$, and when calculated at fixed displacement field, they are known as the Callen charges $\lb Z^{\rm C}\rb$ \cite{callen1949electric}.
Mixed electrostatic boundary conditions $\{\Ef_{\parallel},\Df_{\perp}\}=0$ yield a dynamical charge of mixed Born and Callen type $\lb Z^{\rm 2D}\rb$, although currently no widely available DFT code is capable of performing DFPT calculations with mixed electrostatic boundary conditions. The mixed dynamical charges can be related to the Born charges through the relation between the polarization with different electrostatic boundary conditions:
\beq{eq:P-mix}
\begin{split}
\P^{\rm 2D}_{\parallel} &= \P_{\parallel} \\
P^{\rm 2D}_{\perp} &= \frac{1}{\ep^{\infty}_{zz}} P_{\perp}
\end{split}
\eec
where $\P$ is the polarization calculated for $\{\Ef_{\parallel},\Ef_{\perp}\}=0$ and $\P^{\rm 2D}$ is the polarization calculated for $\{\Ef_{\parallel},\Df_{\perp}\}=0$. Combining Eqs.~\eqref{eq:P-int} and \eqref{eq:P-mix} yields
\beq{eq:Z-mix}
\begin{split}
Z^{\rm 2D}_{\k,\a\b} &= Z^{*}_{\k,\a\b}, \quad \a=x,y \\
Z^{\rm 2D}_{\k,z\b} &= \frac{1}{\ep^{\infty}_{zz}}Z^{*}_{\k,z\b}
\end{split}
\eec
which is similar to the relation between Born charges and Callen charges \cite{ghosez1998dynamical}, however only the third column of $Z^{*}_{\k}$ is modified. 
Eq.~\eqref{eq:Z-mix} implies that, under mixed electrostatic boundary conditions, the dynamical charges naturally asymmetric, provided they are non-diagonal.

The eigenvalues and eigenvectors of the dynamical charges were calculated as a function of relative stacking, and are shown for the B atom in the bottom layer in Fig.~\ref{Fig2}, with the rest available in Ref.~\cite{SM}. 
The in-plane eigenvalues are close to the formal charges, and the out-of-plane one is significantly smaller, due to the 2D nature of the system. 
Thus, the dynamical charges can be visualized as flat discs, see Fig.~\ref{Fig2} (a). 
In general, one of the eigenvectors, $\v_1$, points along the unit cell diagonal $\av_1+\av_2$ ($\av_1$ and $\av_2$ are the lattice vectors of hBN), which is the direction of atomic motion, and the other, $\v_2$, is orthogonal to $\v_1$ and in-plane. 
The third eigenvector, $\v_3$, points out of the plane, see Fig.~\ref{Fig2} (b).
Because of the interlayer electronic charge transfer which occurs with changes in relative stacking, the eigenvalues modulate as one layer slides over the other, see Figs.~\ref{Fig2} (c) and (d).
Additionally, the eigenvectors are non-orthogonal for all except the high-symmetry stackings AA, AB and BA, see Fig.~\ref{Fig2} (e).
As one layer slides over the other, the eigenvectors $\v_1$ and $\v_3$, which are confined to a plane which is normal to the bilayer and along the direction of motion, deflect by $\p_1$ and $\p_3$, respectively, while $\p_2$, which is orthogonal to this plane, is zero for every stacking.
Thus the angle $\th_{13}$, where ${\th_{ij} = \cos^{-1}{\lb \v_i \cdot \v_j\rb}}$, deflects from $\frac{\pi}{2}$ as a function of stacking.

The dynamical charges were decomposed into symmetric and antisymmetric parts: ${Z^{\rm 2D} = \ZS + \ZA}$, where ${Z^{\rm S,A}_{\k,\a\b} = \frac{1}{2} (Z^{\rm 2D}_{\k,\a\b} \pm Z^{\rm 2D}_{\k,\b\a})}$.
The Frobenius norm of the dynamical charges, $\|Z\|_{\rm F}  = \tr{\lb ZZ^{\dagger}\rb}^{\frac{1}{2}}$, which is the analogue of the Euclidean norm of a vector, were calculated, with respect to the AA stacking, and are shown in Fig.~\ref{Fig2} (f).
Although $\ZA$ is traceless, and $\|\ZA\|_{\rm F}$ is small compared to $\|Z^{\rm 2D}\|_{\rm F}$, their spatial modulations are comparable.
Additionally, we can see that there is a direct correspondence between the change in the angle between the eigenvectors and the spatial modulation of $\ZA$.

The eigenvalues of $\ZA$ are always of the form $\lb +i\la^{\rm A}, -i\la^{\rm A}, 0\rb$, with only a single independent parameter 
$\la^{\rm A}$, which is remarkably close to $\D \th_{13}$, although a general relation does not exist \cite{SM}.
The in-plane eigenvectors of $\ZA$, $\v^{\rm A}_{1,2}$ lie along $\av_3 + i( \av_1 + \av_2)$ and are antiparallel, and the out-of-plane eigenvector is $\v^{\rm A}_{3} = \av_1 - \av_2$.
All three eigenvectors of $\ZA$ are constant as a function of stacking.

In general, the dynamical charges can be characterized by
\beq{}
\sum_{i=1}^3 \la_i x_i^2 + \sum_{i \neq j} \la_i \cos(\th_{ij}) x_i x_j \leq 1
\eeq
where $ \cos(\th_{ij}) = \v_i \cdot \v_j$. 
When $\D\th_{ij} \sim \la^{\rm A} \neq 0$, the charges deform from an ideal ellipsoid.

The dynamical charges allow the total polarization to be decomposed into contributions from the motion of different atoms in different directions, 
and additionally, into symmetric and antisymmetric contributions, see Figs.~\ref{Fig2} (g) and (h).
We can see that the total polarization arises mostly from $\ZS$.
Decomposing the total polarization into in-plane sliding and out-of-plane corrugation from the layers, see Figs.~\ref{Fig3} (a) and (b), we can see that most of the polarization is generated from the in-plane sliding of the layers, with higher-order contributions from out-of-plane corrugation.
The directional contributions to the antisymmetric part of the polarization is completely perpendicular, see Figs.~\ref{Fig3} (c) and (d).
While the antisymmetric contributions on the individual B and N atoms are quite large \cite{SM}, they are nearly equal and opposite (apart from the small charge transfer), and thus $\ZA$ makes a negligible contribution to the total polarization as a function of stacking.

\section{Discussion and conclusions} 
In this work, we examine the unconventional nature of ferroelectricity in vdW systems and provide an intuitive description using the dynamical charges.
The dynamical charges themselves are also quite unconventional, where both the eigenvalues and eigenvectors change as a function of relative stacking.
We propose a physical interpretation of the dynamical charges in vdW ferroelectrics as non-ideal ellipsoids which deform as function of relative stacking.
The change in the eigenvalues, i.e.~the major axes of the ellipsoid are a result of charge transfer, and are essential for describing ferroelectricity and topological polarization in vdW materials.
The change in the angles between the eigenvectors, i.e.~the deformation from a regular ellipsoid, denote an anomalous polar response, which is described by the antisymmetric part.
This is a generalization of the physical description of dynamical charges as regular ellipsoids in conventional ferroelectrics, i.e.~ those with a high degree of symmetry, where $\ZA$ is typically zero or negligible.
This visualization is applicable to any anisotropic insulator.
Furthermore, we show that, upon considering the appropriate electrostatic boundary conditions, every 2D system has asymmetric dynamical charges, provided the Born effective charges are non-diagonal.

We note that while in a commensurate bilayer, most of the stacking configurations are dynamically unstable (all except for AB and BA), in a moir\'e superlattice, i.e.~a system with relative strains and/or twists between constituent layers, these stackings are stabilized as they are pinned as dislocations via stacking-engineering.
Thus, the asymmetry of the dynamical charges of vdW ferroelectrics should be realizable by introducing relative twists or lattice mismatches between the layers.
Furthermore, calculating the polarization in terms of the effective charges allows us to understand the origin of the polarization, which mostly arises from the in-plane stacking (shear modes), with higher-order contributions from the out-of-plane corrugation of the layers.

The spatial modulations of $\ZA$ as a function of stacking are comparable to the spatial modulations of $Z^{\rm 2D}$.
Despite this, $\ZA$ does not contribute much to the total polarization in hBN as $\ZA$ fluctuates about zero whereas $Z^{\rm 2D}$ fluctuates about a finite value; 
the nearly equal and opposite antisymmetric contributions arising from the displacements of the B and N atoms almost cancel, resulting in a very small contribution to the total polarization, i.e.~only from the electronic charge transfer and not from ionic contributions.
However, the anomalous dynamical charges in vdW ferroelectrics may result in interesting phenomena in other higher-order material responses.
For example, the relaxed-ion contributions to many response properties are related to the dynamical charges, including permittivity, electromechanical responses such as piezoelectricity and electrostriction \cite{bennett2022generalized}, as well as electro-optic responses and Raman intensities \cite{veithen2004first,veithen2005nonlinear}.
Asymmetric dynamical charges may lead to anomalous responses in such quantities, in particular in the vicinity of a phase transition.

The authors thank X.~He for helpful discussions.
D.B.~acknowledges funding from the University of Li{\`e}ge under special funds for research (IPD-STEMA fellowship programme), the US Army Research Office (ARO) MURI project under grant No.~W911NF-21-0147 and from the Simons Foundation award No.~896626.


%

\clearpage



\section*{Supplementary material}

# Asymmetric dynamical charges in two-dimensional ferroelectrics

Daniel Bennett[1,2,∗] and Philippe Ghosez[1]

[1]*Theoretical Materials Physics, Q-MAT, University of Liège, B-4000 Sart-Tilman, Belgium*

[2]*John A. Paulson School of Engineering and Applied Sciences, Harvard University, Cambridge, Massachusetts 02138, USA*

### Mixed electrostatic boundary conditions

The polarization as a function of relative stacking in bilayer hBN for different electrostatic boundary conditions is shown in Fig. S1.

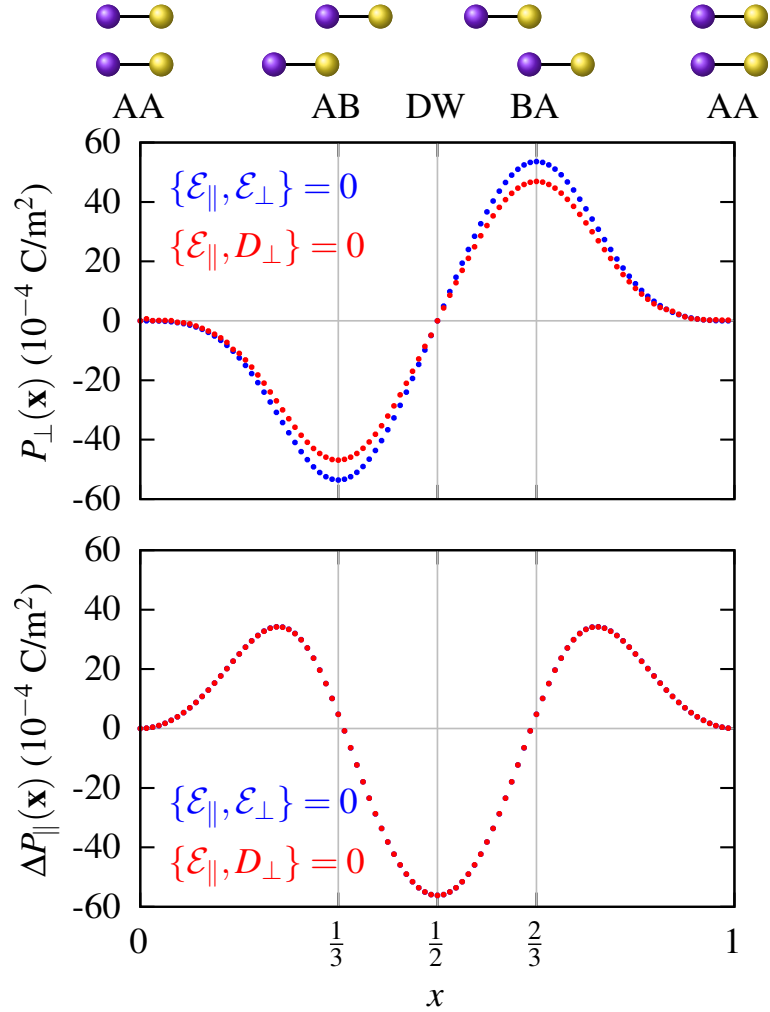

FIG. S1. Polarization as a function of relative stacking in bilayer hBN for different electrostatic boundary conditions: $\{\mathcal{E}_\parallel, \mathcal{E}_\perp\} = 0$ (blue) and $\{\mathcal{E}_\parallel, \mathcal{D}_\perp\} = 0$ (red).

∗ dbennett@seas.harvard.edu

## Eigenvalues and eigenvectors

The eigenvalues of the dynamical charges in hBN are shown as a function of relative stacking in Fig. **??**. The angles $\phi_i$ between between the eigenvectors $\mathbf{v}_i$ and the principle axes are shown in Fig. S2 for each atom, and as a function of relative stacking. The angles $\theta_{ij}$ between between the eigenvectors $\mathbf{v}_i$ and the $\mathbf{v}_j$ are shown in Fig. S3. The imaginary part of $\lambda^A$, the eigenvalue of $Z^A$, multiplied by $\mathrm{rad} = \frac{180}{\pi}$, is shown in Fig. S4 for each atom in bilayer hBN and as a function of relative stacking.

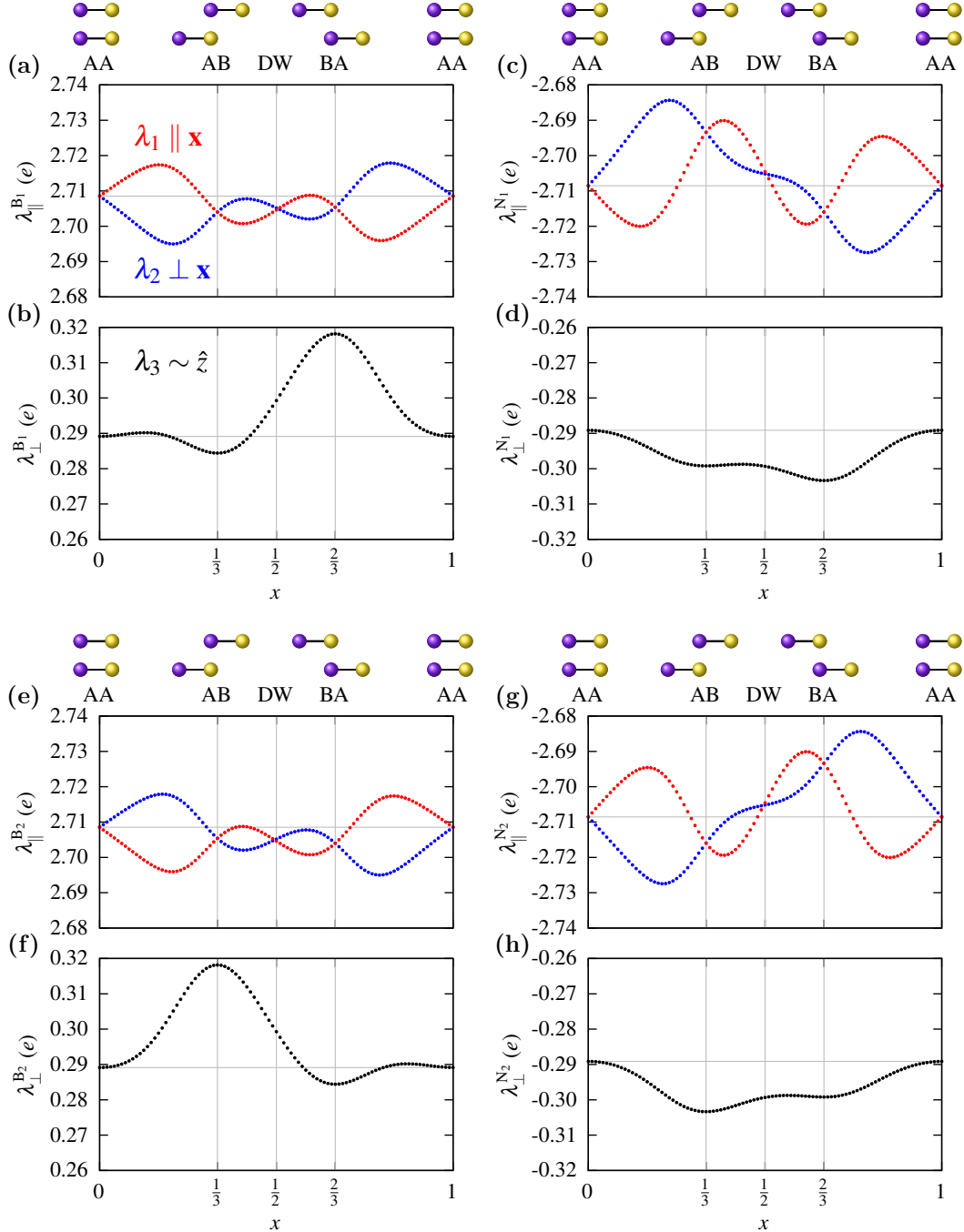

FIG. S2. Eigenvalues of the dynamical charges of each atom in bilayer hBN as a function of relative stacking.

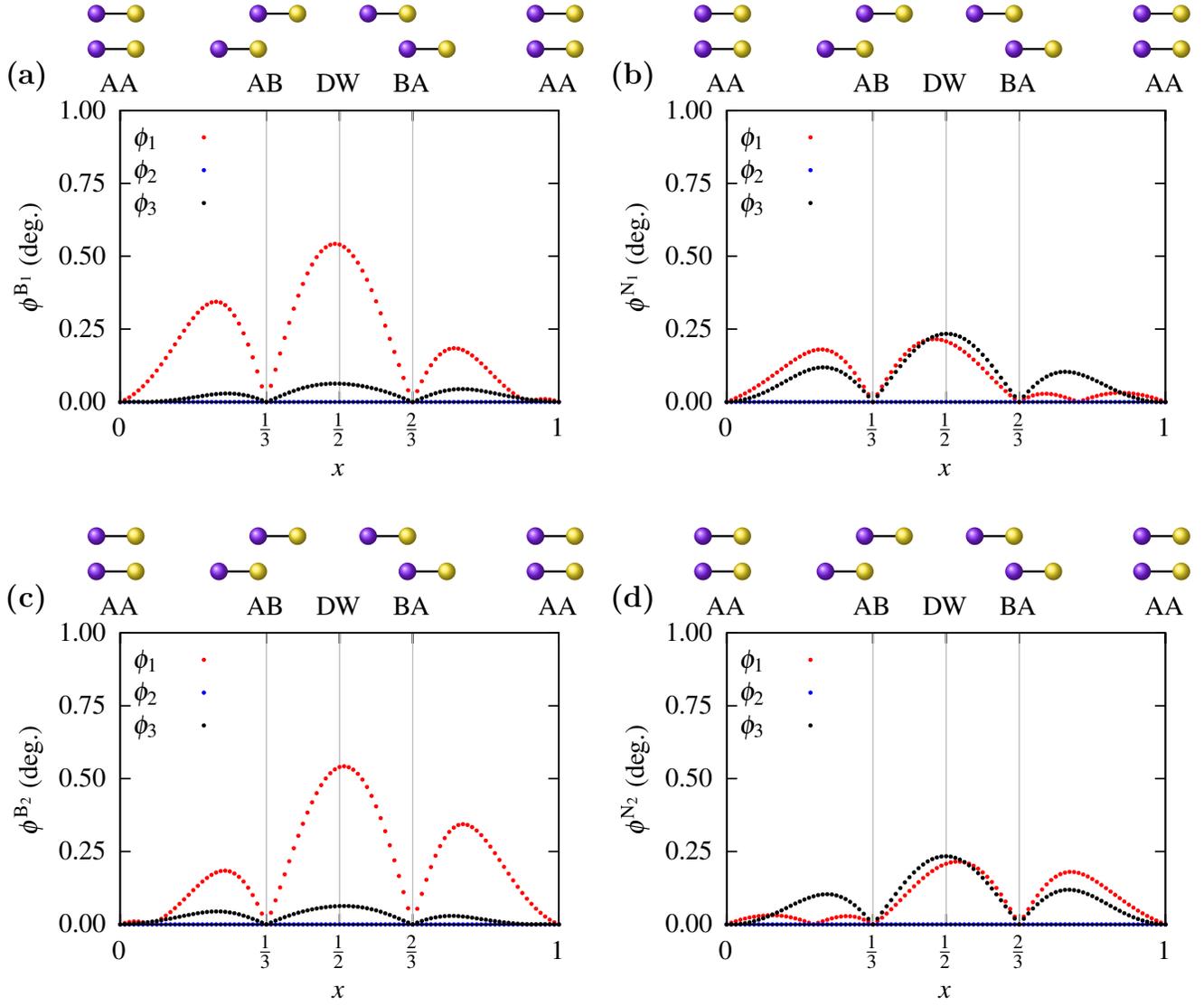

FIG. S3. Angles $\phi_i$ between the eigenvectors $\mathbf{v}_i$ and the principle axes, for each atom in bilayer hBN, as a function of relative stacking.

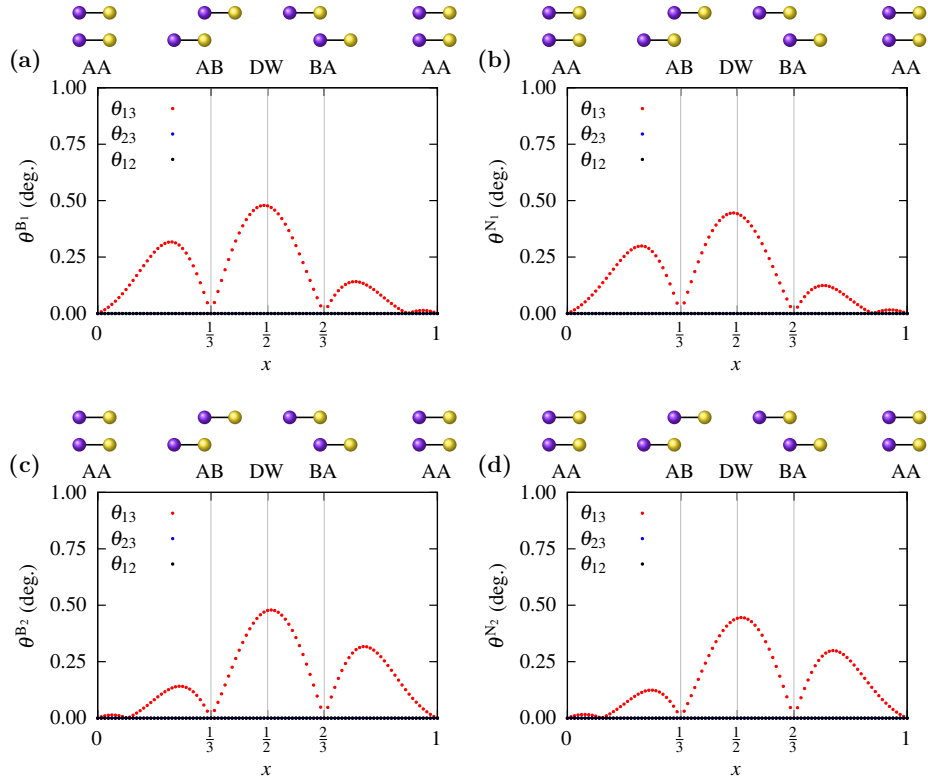

FIG. S4. Angles $\theta_{ij}$ between the eigenvectors $\mathbf{v}_i$ and $\mathbf{v}_j$, for each atom in bilayer hBN, as a function of relative stacking

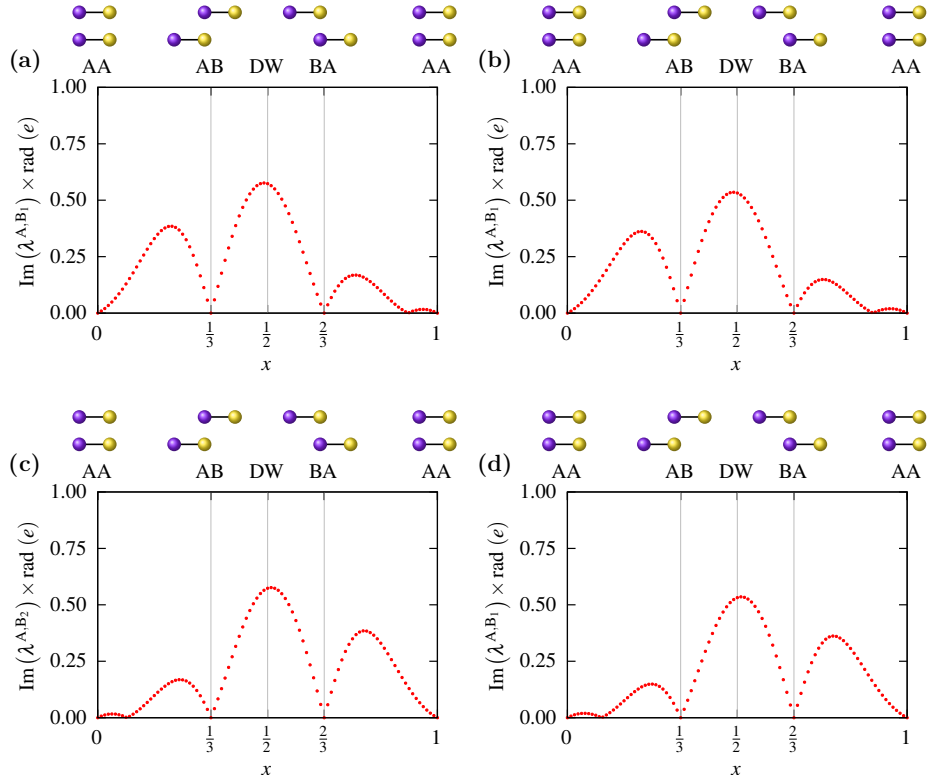

FIG. S5. Imaginary part of $\lambda^A$, the eigenvalue of $Z^A$, multiplied by rad $= \frac{180}{\pi}$, for each atom in bilayer hBN and as a function of relative stacking.

## ANTISYMMETRIC CONTRIBUTION

The norms of the change in total, symmetric and antisymmetric dynamical charge, with respect to the nonpolar AA stacking, are shown in Fig. S5, for each atom and as a function of stacking.

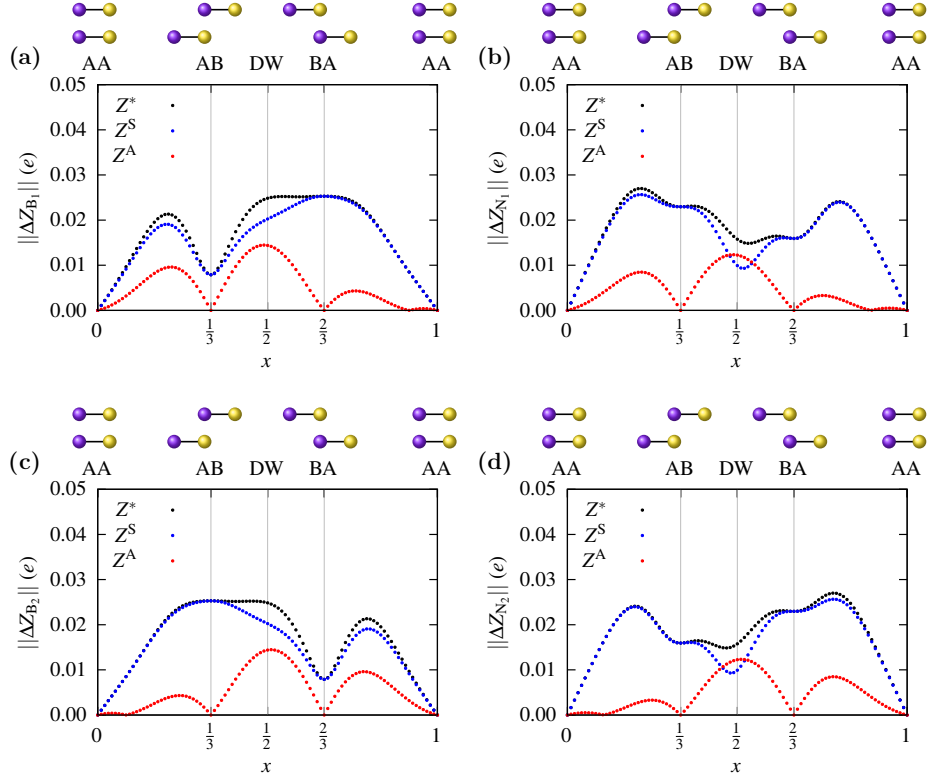

FIG. S6. Norm of the change in total (black), symmetric (red) and antisymmetric (blue) dynamical charge as a function of $\mathbf{x}$: $\Delta||Z|| = ||Z(\mathbf{x}) - Z(0)||$. The modulation in the antisymmetric part is comparable to the modulation of the symmetric part, and is larger for the N atoms for the DW stacking.



\end{document}